\documentclass[11pt,twoside]{article}
\usepackage{asp2006}
\usepackage{psfig}
\usepackage{epsf}
\usepackage{graphics}
\usepackage{lscape}
\def\edcomment#1{\iffalse\marginpar{\raggedright\sl#1\/}\else\relax\fi}
\reversemarginpar

\begin{document}
\title{Properties of 16 Sunspots Observed with Hinode Solar Optical Telescope}
\author{Debi Prasad Choudhary \& Gordon A. MacDonald}
\affil{Department of Physics and Astronomy, California State University, Northridge
18111 Nordhoff St., Northridge Ca, USA, 91330}
\author{Shimizu Toshifumi}
\affil{Institute of Space and Astronautical Science, Japan Aerospace Exploration Agency (ISAS/JAXA)
   3-1-1 Yoshinodai, Sagamihara, Kanagawa 229-8510, Japan}

\begin{abstract}
We studied 16 sunspots with different sizes and shapes using the observations with the Hinode Solar Optical Telescope. The ratio of G-band and CaII H images reveal rich structures both within the umbra and penumbra of most spots. The striking features are the compact blob at the foot point of the umbra side of the penumbral fibrils with disk center-limb side asymmetry. In this paper, we present properties of these features using the spectropolarimetry and images in G-band, CaII and blue filters. We discuss the results using the contemporary models of the sunspots. 
\end{abstract}
\vspace{-0.5cm}

\section{Introduction}

\noindent
{\it "Clearly, observations must point the way, because it is evident now that the sunspot is too complicated a structure to be the product of a single overwhelming theoretical effect. The sunspot results from the conjunction of several effects, and we can spend a lot of time guessing what those effects might be without getting anywhere".~~~~~~~~~~~~Eugene Parker (Personal communication 2010)}

\vspace{0.1 cm}

We analyzed the sunspots observations obtained with Solar Optical Telescope (SOT) onboard Hinode to look for the signatures that are relevant to identify theoretical effect of different models. A sunspot may be made up of individual flux tubes pressed together or a monolithic block of magnetized plasma immersed in the solar convective zone \citep{par79a,mey77}. High spatial resolution observations of sunspot formation show that individual magnetic knots appear on the solar surface that get collected to form larger bodies perhaps by mutual hydrodynamic attraction \citep{par78}. The question still remains that after the  knots are collected together to form a sunspot, do they still retain their individuality? In other words, are there gaps between the individual flux tubes that contain plasma devoid of magnetic field? Existence of field free plasma below the visible layer of the sunspots distinguish the two types of sunspots models \citep{del01}. The physical processes leading to the stability and decay of the sunspot would depend largely on its subphotospheric structure. The time series of high spatial resolution Hinode observations of sunspots, without the interruption of atmospheric seeing, provides opportunity to study the finer structures and their temporal evolution governed by the subphotospheric dynamics. The G-band images serve as a flux tube diagnostic tool due to their association with magnetic concentrations, with efficient heat transport leading to the weakening of molecular band and shifting of optical depth \citep{ish07,san01}. The Ca II H images provides information on the layer slightly higher than the height of G-band formation, the combined understanding of which would provide new understanding of sunspot magnetic structure \citep{car07}.  


\section{Data Analysis}

We use the Broadband Filter Imager (BFI) and Spectropolarimeter (SP) observations obtained with SOT \citep{tsu08}. After initial processing of the level0 data, flux calibration was performed using the measurements of solar radiance through filters made by the detectors on ground and space \citep{shi07}. The ground based measurements were performed by recording the the full field of view images solar disk center by the CCD. The space measurements are made by using the synoptic images that measure the intensities at the solar disk center. The comparison of ground and space  measurements show consistent results. The ratio of the quiet region and the zero air mass solar flux was used to convert sunspot images into solar flux units. The localized bright features in the sunspot that could be successfully fitted with a two dimensional Gaussian function were selected from the co-aligned flux calibrated images. The flux contained in the half width of the Gaussian surface was measured as the intensity of the bright points. The magnetic filling factor was obtained by using inversion codes to the SP observations.  

\section{Discussion and Conclusions}

Figure 1 (a) shows the example of the ratio of G-band and CaII H images of a sunspot observed on 02 May 2007. The important features of the ratio images are: (1) The inner penumbra of the ratio images are brighter compared to the outer part where the CaII H signal is higher compared to the G-band. There is a sharp boundary at the middle of penumbra dividing bright inner and dark outer area. (2) The foot points of the penumbra, which are the locations of the peripheral umbral dots, are brighter compared to the inner umbral dots. The disk-centerward penumbral boundary has brighter points than their limb side counterparts. (3) The umbral dots are not uniformly distributed in the umbra. (4) The umbral dots, closer to the penumbra, are associated with elongated structures directed inwards. (5) Most umbral dots are not circular but elongated elliptical shape. The light bridges consists of compact bright points that have elongated structures. Figure 1 (b) shows the G-band flux and magnetic filling factor as a function of continuum intensity in BFI red filter through at 6685 \AA\/ and 6302.5 \AA\/. The solid curve represent the quadratic function of continuum intensity that follows the trend of both G-band flux and filling factor.   

We interpret these observations using the theoretical models for sunspot structure below and above the visible layer with forest of field free gaps between the flux tubes \citep{par79b,spr06}. Figure 2 describes the physical processes leading to the observed sunspot bright point. The figure shows seven flux tubes in a bundle embedded in field free plasma expanding above photosphere as the gas pressure drops. The field free plasma penetrates in the gap between the flux tubes as shown by gray arrows that transport heat from non magnetic convective zone. The penetrating field free plasma between the flux tubes oscillates, represented by hatched columns with double arrow, due to longitudinal overstability \citep{par79b} and heats the upper layers represented by shaded columns, dissociate the molecules in them and produce umbral dots by altering the optical depth to expose the deeper layers. The inclinaed flux tubes result viewing of larger part of hot flux tube wall \citep{rut01}. The oscillating plasma columns themselves do not appear at the visible surface as blocked by the overlying magnetic canopy. The penumbral flux tubes that are at the periphery of the sunspot get extra heating by direct contact with the convective zone hot plasma represented by hatched side arrows leading to the brighter penumbral foot points. The penumbral flux tubes are highly inclined which contribute to the G-band brightness with limb-disk side asymmetry. As the matter in the flux tube gets heated it expands leading to downward drop in pressure resulting directional flow indicated by arrows in flux tube 1 and 7 \citep{deg93}. The umbra-ward motion of penumbral grains may also be due to this effect. In some locations of the umbra the flux tubes are so dense that there is not much gap between them for the oscillating field free plasma represented by shaded flux bundle between the 4 and 5 flux tube. At these locations there are few umbral dots leading to non uniform distribution in the umbra. In the penumbra, the field free plasma would heat localized regions and produce the bright points represented by shaded and hatched region near flux tube 1 and 7 and produce hotter flux tube plasma in the outer penumbra where the plasma is tenuous due to the expanding flux tubes.

\section{Acknowledgements}
This work was supported by NSF grant ATM-0548260 at California State University Northridge. The data was obtained using Hinode Solar Optical Telescope. Hinode is a Japanese mission developed and launched by ISAS/JAXA, with NAOJ as domestic partner and NASA and STFC (UK) as international partners. It is operated by them in co-operation with ESA and NSC (Norway).

\clearpage

\begin{figure}
\plottwo{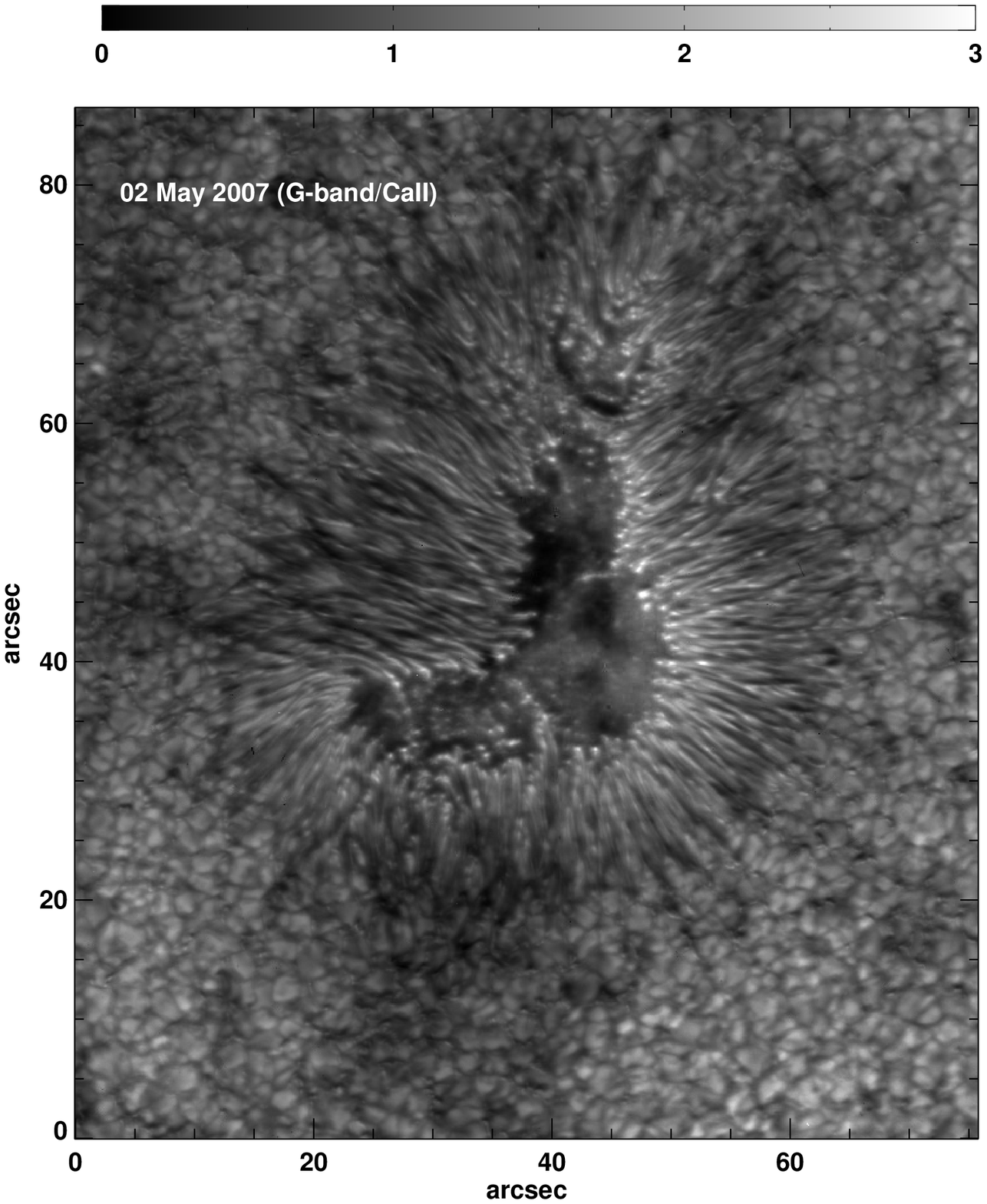}{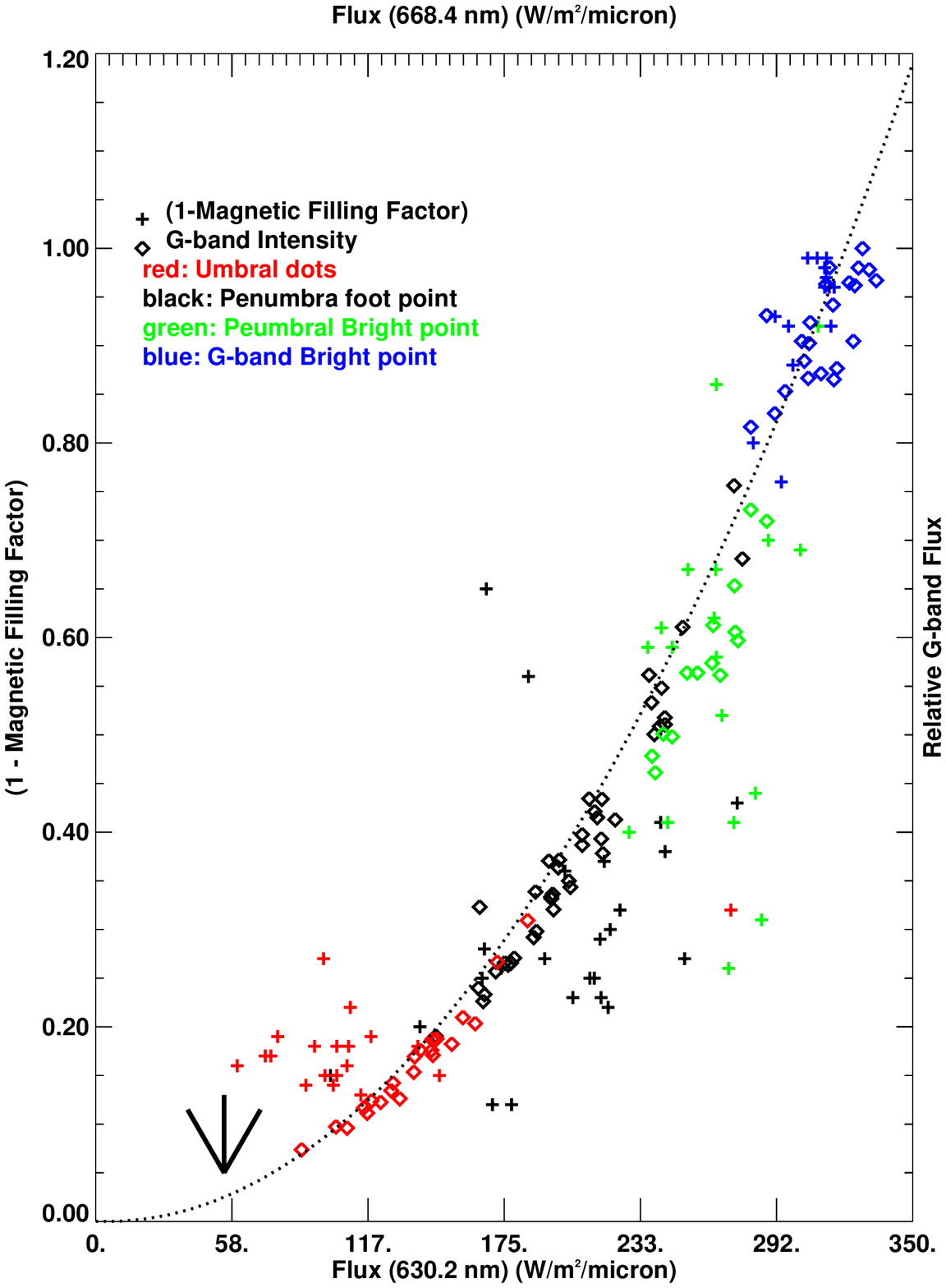}
\caption{(Right) (a) Ratio of G-band and CaII image of the sunspot. The suncenter is to the right. The umbral dots are seen enhanced. The disk center is towards right of the sunspot. (Left) (b) Magnetic Filling factor and G-band intensity of sunspot bright points. The left and bottom axis labels is for the magnetic filling factor. The Top and Right axis labels are for the G-band intensity. The bottom arrow indicates that the actual value of the magnetic filling factor of the umbral bright point may be higher.\label{fig1}}
\end{figure}

\begin{figure}
\plotone{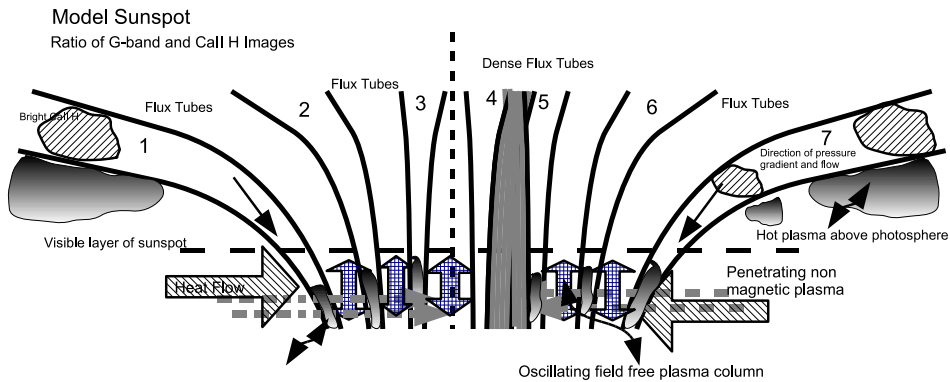}
\caption{ The cartoon of a sunspot that shows theoretical effects of model umbra and penumbra.\label{fig2}}
\end{figure}


\begin{thebibliography}{}
\bibitem[Carlsson et al.(2007)]{car07} 
Carlsson, M. et al.   2007, Pub. Astron. Soc. Japan,
    59, S663
\bibitem[Degenhardt \& Lites(1993)]{deg93} 
Degenhardt, D. \& Lites, B. W.  1993, ApJ,
    416, 875
\bibitem[del Toro Iniesta (2001)]{del01}
del Toro Iniesta, J. C. ASP Conference Series, 248, 35
\bibitem[Ishikawa et al.(2007)]{ish07}
Ishikawa, R., Tsuneta, S., Kitakoshi, Y., Katsukawa, et al., A\&A, 472, 911
\bibitem[Meyer, Schmidt \& Weiss (1977)]{mey77}
Meyer, F., Schmidt, H. U. \& Weiss, N. O. 1977, Mon. Not. Roy. Astr. Soc., 179, 741
\bibitem[Parker(1978)]{par78} 
Parker, E. N.  1978, ApJ, 222, 357
\bibitem[Parker(1979)]{par79a} 
Parker, E. N.  1979, ApJ, 230, 905
\bibitem[Parker(1979)]{par79b} 
Parker, E. N. 1979, ApJ, 234, 333
\bibitem[Rutten et al.(2001)]{rut01} Rutten, R. J. et al.  2001, Astron. Soc. Pac. Conf. Series. 236, 445
\bibitem[Sanches Almeida(2001)]{san01}
Sanches Almeida, J. et al. 2001, ApJ, 555, 978 
\bibitem[Shimizu et al.(2007)]{shi07}
Shimizu, T, Kubo et al., ASP Conference Proceedings, 51.
\bibitem[Spruit \& Scharmer(2006)]{spr06} Spruit, H. C. \& Scharmer, G. B. 2006, A\&A,
    447, 343
\bibitem[Tsuneta et al.(2008)]{tsu08}
Tsuneta, S., Ichimoto, K., Katsukawa, Y., et al, 2008, Sol. Phys., 249, 113.
\end{thebibliography}
\end{document}